\documentclass[sigplan,screen]{acmart}

\usepackage{amsthm}
\newtheorem*{remark}{Agda-ism}

\usepackage{agda}
\usepackage[utf8]{inputenc}
\usepackage{newunicodechar}
\usepackage{newtxmath}
\newunicodechar{⌞}{\ensuremath{\llcorner}}
\newunicodechar{⌟}{\ensuremath{\lrcorner}}
\newunicodechar{′}{\ensuremath{\prime}}
\newunicodechar{−}{\ensuremath{-}}
\newunicodechar{─}{\ensuremath{-}}
\newunicodechar{◆}{\ensuremath{\Diamondblack}}
\newunicodechar{⧫}{\ensuremath{\blacklozenge}}
\newunicodechar{∷}{\ensuremath{::}}
\newunicodechar{∙}{\ensuremath{\bullet}}
\newunicodechar{□}{\ensuremath{\Box}}
\newunicodechar{∎}{\ensuremath{\blacksquare}}
\newunicodechar{⋆}{\ensuremath{\star}}
\newunicodechar{∣}{\ensuremath{|}}

\newunicodechar{₀}{\ensuremath{_0}}
\newunicodechar{₁}{\ensuremath{_1}}
\newunicodechar{₂}{\ensuremath{_2}}
\newunicodechar{₃}{\ensuremath{_3}}

\newunicodechar{ₑ}{\ensuremath{_e}}
\newunicodechar{ᵢ}{\ensuremath{_i}}
\newunicodechar{ₖ}{\ensuremath{_k}}
\newunicodechar{ₘ}{\ensuremath{_m}}
\newunicodechar{ₙ}{\ensuremath{_n}}
\newunicodechar{ᵣ}{\ensuremath{_r}}
\newunicodechar{ₛ}{\ensuremath{_s}}

\newunicodechar{₊}{\ensuremath{_+}}

\newunicodechar{⁺}{\ensuremath{\textsuperscript{+}}}
\newunicodechar{⁻}{\ensuremath{\textsuperscript{-}}}

\newunicodechar{²}{\ensuremath{^2}}

\newunicodechar{ᵈ}{\ensuremath{^d}}
\newunicodechar{ⁱ}{\ensuremath{^i}}
\newunicodechar{ˡ}{\ensuremath{^l}}
\newunicodechar{ʳ}{\ensuremath{^r}}
\newunicodechar{ˢ}{\ensuremath{^s}}

\newunicodechar{ᴬ}{\ensuremath{^A}}
\newunicodechar{ᴮ}{\ensuremath{^B}}
\newunicodechar{ᴵ}{\ensuremath{^I}}
\newunicodechar{ᴿ}{\ensuremath{^R}}
\newunicodechar{ᵀ}{\ensuremath{^T}}
\newunicodechar{ᵁ}{\ensuremath{^U}}
\newunicodechar{ⱽ}{\ensuremath{^V}}

\newunicodechar{⋯}{\ensuremath{\cdots}}
\newunicodechar{∶}{\ensuremath{:}}

\newunicodechar{∼}{\ensuremath{\sim}}
\newunicodechar{≡}{\ensuremath{\equiv}}
\newunicodechar{≢}{\ensuremath{\not\equiv}}
\newunicodechar{≟}{\mbox{\tiny\ensuremath{\stackrel{?}{=}}}}
\newunicodechar{≈}{\ensuremath{\approx}}

\newunicodechar{≤}{\ensuremath{\le}}
\newunicodechar{≥}{\ensuremath{\ge}}

\newunicodechar{↦}{\ensuremath{\mapsto}}
\newunicodechar{⇑}{\ensuremath{\Uparrow}}
\newunicodechar{→}{\ensuremath{\rightarrow}}
\newunicodechar{←}{\ensuremath{\leftarrow}}
\newunicodechar{⇒}{\ensuremath{\Rightarrow}}
\newunicodechar{⇉}{\ensuremath{\rightrightarrows}}

\newunicodechar{∂}{\ensuremath{\partial}}
\newunicodechar{∋}{\ensuremath{\ni}}
\newunicodechar{∞}{\ensuremath{\infty}}
\newunicodechar{∀}{\ensuremath{\forall}}
\newunicodechar{∃}{\ensuremath{\exists}}
\newunicodechar{⊢}{\ensuremath{\vdash}}
\newunicodechar{⟨}{\ensuremath{\langle}}
\newunicodechar{⟩}{\ensuremath{\rangle}}
\newunicodechar{⊤}{\ensuremath{\top}}
\newunicodechar{∘}{\ensuremath{\circ}}
\newunicodechar{⊎}{\ensuremath{\uplus}}
\newunicodechar{×}{\ensuremath{\times}}
\newunicodechar{ℕ}{\ensuremath{\mathbb{N}}}
\newunicodechar{⟦}{\ensuremath{\llbracket}}
\newunicodechar{⟧}{\ensuremath{\rrbracket}}
\newunicodechar{∈}{\ensuremath{\in}}
\newunicodechar{↑}{\ensuremath{\uparrow}}
\newunicodechar{¬}{\ensuremath{\neg}}
\newunicodechar{⊥}{\ensuremath{\bot}}
\newunicodechar{↝}{\ensuremath{\leadsto}}
\newunicodechar{↶}{\ensuremath{\curvearrowleft}}
\newunicodechar{↺}{\ensuremath{\circlearrowleft}}
\newunicodechar{⊔}{\ensuremath{\sqcup}}
\newunicodechar{⨆}{\ensuremath{\bigsqcup}}
\newunicodechar{∩}{\ensuremath{\cap}}
\newunicodechar{∪}{\ensuremath{\cup}}

\newunicodechar{ℓ}{\ensuremath{\ell}}

\newunicodechar{Δ}{\ensuremath{\Delta}}
\newunicodechar{Γ}{\ensuremath{\Gamma}}
\newunicodechar{Σ}{\ensuremath{\Sigma}}
\newunicodechar{Θ}{\ensuremath{\Theta}}
\newunicodechar{Ω}{\ensuremath{\Omega}}

\newunicodechar{α}{\ensuremath{\alpha}}
\newunicodechar{β}{\ensuremath{\beta}}
\newunicodechar{δ}{\ensuremath{\delta}}
\newunicodechar{ε}{\ensuremath{\varepsilon}}
\newunicodechar{φ}{\ensuremath{\phi}}
\newunicodechar{γ}{\ensuremath{\gamma}}
\newunicodechar{ι}{\ensuremath{\iota}}
\newunicodechar{κ}{\ensuremath{\kappa}}
\newunicodechar{λ}{\ensuremath{\lambda}}
\newunicodechar{μ}{\ensuremath{\mu}}
\newunicodechar{ψ}{\ensuremath{\psi}}
\newunicodechar{η}{\ensuremath{\eta}}
\newunicodechar{ρ}{\ensuremath{\rho}}
\newunicodechar{σ}{\ensuremath{\sigma}}
\newunicodechar{τ}{\ensuremath{\tau}}
\newunicodechar{ξ}{\ensuremath{\xi}}
\newunicodechar{ζ}{\ensuremath{\zeta}}
\newunicodechar{Π}{\ensuremath{\Pi}}

\newunicodechar{𝓒}{\ensuremath{\mathcal{C}}}
\newunicodechar{𝓔}{\ensuremath{\mathcal{E}}}
\newunicodechar{𝓕}{\ensuremath{\mathcal{F}}}
\newunicodechar{𝓡}{\ensuremath{\mathcal{R}}}
\newunicodechar{𝓢}{\ensuremath{\mathcal{S}}}
\newunicodechar{𝓣}{\ensuremath{\mathcal{T}}}
\newunicodechar{𝓥}{\ensuremath{\mathcal{V}}}
\newunicodechar{𝓦}{\ensuremath{\mathcal{W}}}

\usepackage{catchfilebetweentags}
\input{robust-catch}
\usepackage{todonotes}
\setuptodonotes{inline}

\newcommand{\AK}[1]{\AgdaKeyword{\small#1}}

\newcommand{\AS}[1]{\AgdaSymbol{\small#1}}

\newcommand{\AN}[1]{\AgdaNumber{\small#1}}
\newcommand{\AD}[1]{\AgdaDatatype{\small#1}}
\newcommand{\AF}[1]{\AgdaFunction{\small#1}}
\newcommand{\AR}[1]{\AgdaRecord{\small#1}}
\newcommand{\ARF}[1]{\AgdaField{\small#1}}
\newcommand{\AB}[1]{\AgdaBound{\small#1}}
\newcommand{\AIC}[1]{\AgdaInductiveConstructor{\small#1}}
\newcommand{\AC}[1]{\AgdaComment{\small#1}}

\usepackage{tikz}
\usepackage{circuitikz}
\ctikzset{logic ports=ieee}

\usepackage{microtype}
\DisableLigatures[-]{}

\newtoggle{BLIND}
\togglefalse{BLIND}

\usepackage{hyperref}
\usepackage{cleveref}

\usepackage{mathpartir}

\acmConference[PEPM '24]{Proceedings of the 2024 ACM SIGPLAN International Workshop on Partial Evaluation and Program Manipulation}{January 16, 2024}{London, UK}
\acmBooktitle{Proceedings of the 2024 ACM SIGPLAN International Workshop on Partial Evaluation and Program Manipulation (PEPM '24), January 16, 2024, London, UK}

\begin{document}

%%
%% The "title" command has an optional parameter,
%% allowing the author to define a "short title" to be used in page headers.
\title{Scoped and Typed Staging by Evaluation}

%%
%% The "author" command and its associated commands are used to define
%% the authors and their affiliations.
%% Of note is the shared affiliation of the first two authors, and the
%% "authornote" and "authornotemark" commands
%% used to denote shared contribution to the research.
\author{Guillaume Allais}
\orcid{0000-0002-4091-657X}
\affiliation{%
  \institution{University of Strathclyde}
  \city{Glasgow}
  \country{UK}
}
\email{guillaume.allais@ens-lyon.org}
%%
%% By default, the full list of authors will be used in the page
%% headers. Often, this list is too long, and will overlap
%% other information printed in the page headers. This command allows
%% the author to define a more concise list
%% of authors' names for this purpose.
% \renewcommand{\shortauthors}{}

%%
%% The abstract is a short summary of the work to be presented in the
%% article.
\begin{abstract}
\end{abstract}

%%
%% The code below is generated by the tool at http://dl.acm.org/ccs.cfm.
%% Please copy and paste the code instead of the example below.
%%


%%
%% Keywords. The author(s) should pick words that accurately describe
%% the work being presented. Separate the keywords with commas.
%% A "teaser" image appears between the author and affiliation
%% information and the body of the document, and typically spans the
%% page.
%% \begin{teaserfigure}
%%   \includegraphics[width=\textwidth]{sampleteaser}
%%   \caption{Seattle Mariners at Spring Training, 2010.}
%%   \Description{Enjoying the baseball game from the third-base
%%   seats. Ichiro Suzuki preparing to bat.}
%%   \label{fig:teaser}
%% \end{teaserfigure}

\received{2023-10-20}
\received[accepted]{2023-11-20}

\begin{abstract}
  Using a dependently typed host language, we give a well
  scoped-and-typed by construction presentation of a minimal
  two level simply typed calculus with a static and a dynamic
  stage.
  The staging function partially evaluating the parts of a term
  that are static is obtained by a model construction inspired
  by normalisation by evaluation.

  We then go on to demonstrate how this minimal language can be
  extended to provide additional metaprogramming capabilities,
  and to define a higher order functional language evaluating
  to digital circuit descriptions.
\end{abstract}

\maketitle

\section{Introduction}

Staged compilation, by running arbitrary programs at compile
time in order to generate code, is a way to offer users
metaprogramming facilities.
Kov{\'{a}}cs demonstrated that the notion of two level
type theories, originally introduced in homotopy theory,
can be repurposed to describe layered languages equipped
with a staging operation partially evaluating the terms
in the upper layer~\cite{DBLP:journals/pacmpl/Kovacs22}.

In order to enable the mechanised study of such systems,
we give an intrinsically scoped-and-typed treatment of
various two level simply typed calculi and their
corresponding staging operations evaluating away all of
the static subterms.
We obtain these staging operations by performing type-directed
model constructions reminiscent of the ones used for normalisation
by evaluation, hence the title of this paper.

We then build a system that takes seriously Kov{\'{a}}cs'
remark that the static and dynamic layers do not need to have
exactly the same features. Its static layer is a higher order
functional language while the dynamic one corresponds to
digital circuit descriptions.
This casts existing work on high level languages for quantum
circuit descriptions into a new light as two level theories.

This work has been fully formalised using Agda~\cite{DBLP:conf/afp/Norell08}
as our host language (but any implementation
of Martin-Löf type theory~\cite{DBLP:books/daglib/0000395}
with inductive families~\cite{DBLP:journals/fac/Dybjer94}
would do).

\section{A Primer on Intrinsically Typed λ-Calculi}\label{sec:intrinsictyping}

Let us start with a quick primer on intrinsically scoped-and-typed λ-calculi
defined in a dependently typed host language. The interested reader
can refer to ACMM~\cite{DBLP:conf/cpp/Allais0MM17} for a more in-depth
presentation of this approach.

\subsection{Object Types and Contexts}

We first give an inductive definition of object types.
We call it \AF{Type} and its own type is \AF{Set},
the type of all small types in Agda.
It has two constructors presented in generalised algebraic
datatype fashion.
We use \AIC{`α} as our base type,
and (\AB{A} \AIC{`⇒} \AB{B}) is the type of functions from
\AB{A} to \AB{B}.

\ExecuteMetaData[STLC.tex]{type}

\begin{remark}[Syntax Highlighting]
  All of the code snippets in this paper are semantically highlighed:
  keywords are \AK{orange}, definitions and types are \AF{blue},
  data constructors are \AIC{green}, bound variables are
  \AB{slanted}, and comments are \AC{brown}.
\end{remark}

\begin{remark}[Implicit Prenex Polymorphism]
  We extensively use Agda's \AK{variable} mechanism: all of
  the seemingly unbound names will in fact have been automatically
  quantified over in a prenex position provided that they
  have been declared beforehand.
\end{remark}

The following block for instance announces that from now
on unbound \AB{A}s, \AB{B}s, and \AB{C}s stand for implicitly bound
\AD{Type} variables.

\ExecuteMetaData[STLC.tex]{typevariables}

Next, we form contexts as left-nested lists of types using
constructor names similar to the ones typically used in
type judgments.
Contexts may be the empty context \AIC{ε}
or a compound context (\AB{Γ} \AIC{,} \AB{A}) obtained
by extending an existing context \AB{Γ} on the right
with a newly bound (nameless) variable of type \AB{A}.

\ExecuteMetaData[STLC.tex]{context}

\subsection{Manipulating Indexed Types}

In this paper we are going to conform to the convention
of only mentioning context \emph{extensions} when
presenting judgements.
That is to say we will write the application and
λ-abstraction rules as they are in the right column
rather than the left one where the ambient context Γ
is explicitly threaded.

\noindent
\begin{minipage}{.2\textwidth}
  \begin{mathpar}
    \inferrule{Γ ⊢ f : A → B \and Γ ⊢ t : A}{Γ ⊢ f \, t : B} \and
    \inferrule{Γ, x : A ⊢ b : B}{Γ ⊢ λx.b : A → B}
  \end{mathpar}
\end{minipage}\hfill
\begin{minipage}{.2\textwidth}
  \begin{mathpar}
    \inferrule{f : A → B \and t : A}{f \, t : B} \and
    \inferrule{x : A ⊢ b : B}{λx.b : A → B}
  \end{mathpar}
\end{minipage}

To do so, we need to introduce a small set of combinators
to manipulate indexed definitions. These are commonplace
and already present in Agda's standard library.
First, \AF{∀[\_]} universally quantifies over its argument's index;
it is meant to be used to surround a complex expression built
up using the other combinators.

\ExecuteMetaData[STLC.tex]{forall}

Second, the suggestively named \AF{\_⊢\_} allows us to modify
the index; it will be useful to extend a context with freshly
bound variables.

\ExecuteMetaData[STLC.tex]{update}

Third, we can form index-respecting function spaces.

\ExecuteMetaData[STLC.tex]{arrow}

Finally, the pointwise lifting of pairing is called \AF{\_∩\_};
it will only come into play in \Cref{sec:stagingmodelprod}.

\ExecuteMetaData[STLC.tex]{product}

We include below an artificial example of a type written using
the combinators together with its full expansion using explicit
context-passing.

\begin{AgdaSuppressSpace}
  \ExecuteMetaData[STLC.tex]{swaptype}
  \ExecuteMetaData[STLC.tex]{swaptypenormalised}
\end{AgdaSuppressSpace}

\subsection{Intrinsically Typed Variables}

Our first inductive family~\cite{DBLP:journals/fac/Dybjer94}
is called \AD{Var} and formalises
what it means for a variable of type \AB{A} to be present in context \AB{Γ}.
It is indexed over said type and context.
We present it side by side
with the corresponding inference rules for the typing judgement
for variables denoted ($\cdot :_v \cdot$).
The first constructor (\AIC{here}) states that in a non-empty context
where the most local variable has type \AB{A} we can indeed obtain a
variable of type \AB{A}.
The second one (\AIC{there}) states that if a variable of type \AB{A}
is present in a context then it also is present in the same context
extended with a freshly bound variable of type \AB{B}.

\ExecuteMetaData[STLC.tex]{var}

\begin{mathpar}
  \inferrule{ }{x : A ⊢ x :_v A} \and
  \inferrule{x :_v A}{y : B ⊢ x :_v A}
\end{mathpar}

This is a standard definition
corresponding to a scoped-and-typed variant of De Bruijn
indices~\cite{de1972lambda,DBLP:journals/scp/BellegardeH94,DBLP:conf/csl/AltenkirchR99,DBLP:journals/jfp/BirdP99}:
\AIC{here} corresponds to zero, and \AIC{there} to successor.

\subsection{Intrinsically Typed Terms}

We are now ready to give the type of intrinsically typed terms.
It is once again an inductive family indexed over a type and a
context; its declaration is as follows.

\ExecuteMetaData[STLC.tex]{termdecl}

We will introduce constructors in turn, each paralleled by its
counterpart as an inference rule.
We start with the variable rule: a variable of type \AB{A}
forms a valid term of type \AB{A}.
As you can see below, we use a line lexed as a comment (\AC{----})
to suggestively typeset the constructor's type like the
corresponding rule.

\noindent
\begin{minipage}[t]{.25\textwidth}
  \ExecuteMetaData[STLC.tex]{termvar}
\end{minipage}\hfill
\begin{minipage}[t]{.17\textwidth}
\begin{mathpar}
  \inferrule{x :_v A}{x : A}
\end{mathpar}
\end{minipage}

Next we have the constructor for applications. It states that
by combining a term whose type is a function type from \AB{A} to \AB{B}
and a term of type \AB{A}, we obtain a term of type \AB{B}.

\noindent
\begin{minipage}[t]{.25\textwidth}
  \ExecuteMetaData[STLC.tex]{termapp}
\end{minipage}\hfill
\begin{minipage}[t]{.17\textwidth}
\begin{mathpar}
  \inferrule{f : A → B \and t : A}{f \, t : B}
\end{mathpar}
\end{minipage}

Last but not least, the rule for λ-abstraction is the only rule
with a premise mentioning a context extension. It states that
we can build a term for a function from \AB{A} to \AB{B} by
building the function's body of type \AB{B} in a context extended
by a freshly bound variable of type \AB{A}.

\noindent
\begin{minipage}[t]{.25\textwidth}
  \ExecuteMetaData[STLC.tex]{termlam}
\end{minipage}\hfill
\begin{minipage}[t]{.17\textwidth}
\begin{mathpar}
  \inferrule{x : A ⊢ b : B}{λx.b : A → B}
\end{mathpar}
\end{minipage}

Putting it all together, we obtain the following inductive family.

\begin{AgdaAlign}
\begin{AgdaSuppressSpace}
  \ExecuteMetaData[STLC.tex]{term}
\end{AgdaSuppressSpace}
\end{AgdaAlign}

This gives us the intrinsically scoped-and-typed syntax for
the simply typed lambda calculus.
And we give our first example: the identity function ($λ.0$ in de Bruijn notation).

\ExecuteMetaData[STLC.tex]{id}

As any well behaved syntax should,
it is stable under weakening as we are going to see shortly.

\subsection{Weakening}

Following Altenkirch, Hofmann, and Streicher~\cite{DBLP:conf/ctcs/AltenkirchHS95}
we start by defining the category of weakenings with contexts
as objects and the following inductive family as morphims.

\ExecuteMetaData[STLC.tex]{ope}

This relation on contexts, also known as order-preserving embeddings in the
literature, is a first order description of order-preserving
injections:
\AIC{done} is the trivial injection of the empty context into
itself;
\AIC{keep} extends an existing injection into one that preserves
the most local variable;
and \AIC{drop} records that the most local variable of the target
context does not have a pre-image via the injection.

We can define identity and composition of these morphisms (we leave
the definitions out but they are available in the accompanying material).

\noindent
\begin{minipage}{.15\textwidth}
  \ExecuteMetaData[STLC.tex]{lerefl}
\end{minipage}\hfill
\begin{minipage}{.27\textwidth}
  \ExecuteMetaData[STLC.tex]{letrans}
\end{minipage}

These order-preserving embeddings all have an action on suitably
well behaved scoped families. We will call these actions weakening
principles, and introduce the following type synonym to describe them.

\ExecuteMetaData[STLC.tex]{weaken}

The action on variables is given
by the following \AF{wkVar} definition. It is defined by
induction over the renaming and case analysis on the de Bruijn
index if the most local variable happens to be in both contexts.

\ExecuteMetaData[STLC.tex]{weakVar}

The action on terms is purely structural, with the caveat that
the weakening needs to be amended when going under a binder: the
most recently bound variable is present in both the source and
target contexts and so we use \AIC{keep} to mark it as retained.

\ExecuteMetaData[STLC.tex]{weakTerm}

Using these results, we can define function composition
as a pseudo constructor: provided $g$ and $f$, we form
$λx.g\,\left(f\, x\right)$ i.e. we use $g$ and $f$ in a context extended
with $x$ hence the need for weakening.

\label{def:composition}
\ExecuteMetaData[STLC.tex]{composition}

\begin{remark}[Lexing of Identifiers]
  Ignoring details about reserved characters for now: any
  space-free string of unicode characters is considered a
  single identifier.
  Correspondingly, in the example above \AB{Γ≤Γ,A} is a single
  identifier whose named is meant to document for the human reader
  what its type looks like: the embedding of a context \AB{Γ} into
  its extension by a single free variable of type \AB{A}.
\end{remark}

\subsection{Normalisation by Evaluation}

It is now time to define an evaluation function for this syntax.
By the end of this section, we will have a function \AF{eval}
turning terms into Kripke-style values,
provided that we have an environment assigning values
to each of the term's free variables. It will have the following type.

\ExecuteMetaData[STLC.tex]{evaldecl}

\subsubsection{Kripke Function Spaces}

This whole process is based on Kripke semantics for intuitionistic
logic~\cite{DBLP:journals/apal/MitchellM91}.
As a consequence one of the central concepts is closure under future
worlds, here context extensions.
This idea is captured by the definition of the \AR{□} record: we
can inhabit (\AR{□} \AB{P} \AB{Γ}) whenever for any extension
\AB{Δ} of \AB{Γ} we are able to construct a proof of (\AB{P} \AB{Δ}).

\ExecuteMetaData[STLC.tex]{box}

For more information on \AR{□} and its properties, see
Allais, Atkey, Chapman, McBride, and
McKinna~\cite[Section~3.1]{DBLP:journals/jfp/AllaisACMM21}.
We will only use the fact that it is a comonad, that is to say that
we can define \AF{extract} and \AF{duplicate} thanks to
the fact that the embedding relation is a preorder.

\ExecuteMetaData[STLC.tex]{extract}
\ExecuteMetaData[STLC.tex]{duplicate}

\begin{remark}[Copattern matching]
The definition of \AF{duplicate} proceeds by
copattern-matching~\cite{DBLP:conf/popl/AbelPTS13}.
This allows us to define values of a record type by defining
the result of taking each of its projections.

In this instance it is particularly useful because the
definition's type involves nested \AR{□} records and
the type of \AR{□}'s projection (\ARF{run□}) is itself
a function type taking an (implicit) context and a
weakening into that context.
This notation allows us to alternate entering the record
and binding extra arguments such as \AB{σ}.
\end{remark}

Kripke function spaces then correspond to functions inside a box,
hence the following definition.

\ExecuteMetaData[STLC.tex]{kripke}

The comonadic structure of \AF{□} additionally ensures we can define
semantic application (\AF{\_\$\$\_}) and weakening of Kripke function
spaces.

%% \noindent
%% \begin{minipage}{0.45\textwidth}
  \ExecuteMetaData[STLC.tex]{semapp}
%% \end{minipage}\hfill
%% \begin{minipage}{0.5\textwidth}
  \ExecuteMetaData[STLC.tex]{weakKripke}
%% \end{minipage}

Finally, we introduce a notation to hide away \AR{□}-related
notions when building \AF{Kripke} functions.
After the following declarations we can write
\AIC{λλ[} \AB{σ} \AIC{,} \AB{v} \AIC{]} \AB{b} to implement a function of type
(\AF{Kripke} \AB{P} \AB{Q} \AB{Γ}).

\ExecuteMetaData[STLC.tex]{mkbox}

\begin{remark}[Syntax Declarations]
  A \AK{syntax} declaration introduces syntactic sugar that is
  allowed to perform variable binding, or take arguments in a
  seemingly out-of-order manner.
  In the above declaration the left hand side describes the
  actual term and the right hand side its newly introduced
  sugared form.
\end{remark}

We now have all of the ingredients necessary to perform the model
construction allowing us to implement a normaliser.

\subsubsection{Model Construction}

This step follows standard techniques for normalisation
by evaluation~\cite{DBLP:conf/lics/BergerS91,DBLP:journals/mscs/CoquandD97,DBLP:journals/lisp/Coquand02}.
The family of values is defined by induction on the value's
type.
Values of a base type are neutral terms (this is not enforced here and
we are happy to simply reuse \AF{Term}) while values of a function type
are Kripke function spaces between values of the domain and values of
the codomain.

\ExecuteMetaData[STLC.tex]{value}

We prove that values can be weakened by using the fact they are
defined in terms of families already known to be amenable to weakenings.

\ExecuteMetaData[STLC.tex]{weakValue}

Environments are functions associating a \AF{Value}
to each \AF{Var}iable in scope.

\ExecuteMetaData[STLC.tex]{env}

In the upcoming definition of the evaluation function, environments
will in general simply be threaded through. They will only need to
be modified when going under a binder. This binder, interpreted as
a Kripke function space, will provide a context weakening and a value
living in that context.
The environment will have to be extended with the value while its
existing content will need to be transported, along the weakening,
into the bigger context.
The \AF{extend} definition combines these two operations into a single
one. It is defined in copattern style:
{\AS{.}\ARF{run□}} builds a box while {\AS{.}\ARF{get}} builds the
returned environment. The definition proceeds by case analysis on the
variable to be mapped to a value: if it is the newly bound one, we
immediately return the value we just obtained, and otherwise we look
up the associated value in the old environment and use \AB{σ} to
appropriately weaken it.

\ExecuteMetaData[STLC.tex]{extend}

The evaluation function maps terms to values provided that
an environment assigns a value to every free variable in scope.
It is defined by induction on the term and maps every construct
to its semantical counterpart: variables become environment lookups,
applications become Kripke applications, and λ-abstractions become
Kripke functions.

\begin{AgdaSuppressSpace}
\ExecuteMetaData[STLC.tex]{eval}
\end{AgdaSuppressSpace}

A typical normalisation by evaluation presentation would
conclude with the definition of a reification function
extracting a term from a value in a type-directed manner
before defining normalisation as the composition of evaluation
and reification.
This last step will however not be useful for our study of two
level calculi and so we leave it out. It can be found in details
in Catarina Coquand's work on normalisation by evaluation for
a simply typed λ-calculus with explicit
substitutions~\cite{DBLP:journals/lisp/Coquand02}.

Now that we have seen how to define a small well scoped-and-typed
language and implement an evaluation function by performing a model
construction, we can start looking at extending it to a two level
language.

\section{Minimal Intrinsically Typed Two Level Type Theory}

We start with the smallest two level calculus we can possibly define
by extending the simply typed λ-calculus as defined in the previous
section with quotes (\AIC{`⟨\_⟩}) and splices (\AIC{`∼\_}).

This will enable us to write and stage simple programs such as the following.

\ExecuteMetaData[BasicTwoTT.tex]{testid01}

The three-place relation (\AB{A} \AF{∋} \AB{s} \AF{↝} \AB{t}) states
that staging a term \AB{s} at type \AB{A} yields the term \AB{t}.
Here, \AF{`idᵈ} is a dynamic identity function
while \AF{`idˢ} is a static one,
\AIC{`⟨\_⟩} quotes a static term inside a dynamic one,
and \AIC{`∼\_} splices a dynamic term in a static one.
Correspondingly, staging will partially evaluate the call to
\AF{`idˢ} as well as all the quotes and splices while leaving
the rest of the term intact.
Hence the result: the call to the static identity function has
fully reduced but the call to the dynamic one has been preserved.

\subsection{Phases, Stages, and Types}

We start by defining a sum type of phases denoting whether
we are currently writing source (\AIC{src}) code or inspecting
staged (\AIC{stg}) code that has already been partially evaluated.

\ExecuteMetaData[BasicTwoTT.tex]{phase}

Additionally, our notion of types is going to be explicitly
indexed by the stage they live in. These stages are themselves
indexed over the phase they are allowed to appear in.
The static (\AIC{sta}) stage is only available in the \AIC{src}
phase: once code has been staged, all of its static parts will
be gone.
The dynamic (\AIC{dyn}) stage however will be available in both
phases, hence the unconstrained index \AB{ph}.

\ExecuteMetaData[BasicTwoTT.tex]{stage}

We can now define our inductive family of simple types indexed
by their stage.

\ExecuteMetaData[BasicTwoTT.tex]{types}

We have both static and dynamic terms of base type,
hence the unconstrained indices \AB{ph} and \AB{st}
for the constructor \AIC{`α}.
The constructor \AIC{‘⇑\_} allows us to embed dynamic
types into static ones; (\AIC{‘⇑} \AB{A}) is effectively
the type of \emph{programs} that will compute a value of
type \AB{A} at runtime. This is only available in the
\AIC{src} phase.
Function types are available in both layers provided that
they are homogeneous: both the domain and codomain need
to live in the same layer.

Purely dynamic types in the source phase have a direct
counterpart in the staged one. We demonstrate this by
implementing the following \AF{asStaged} function.

\ExecuteMetaData[BasicTwoTT.tex]{asStaged}

It is essentially the identity function except for the
fact that its domain and codomain have different indices.

\subsection{Intrinsically Scoped and Typed Syntax}

We skip over the definition of contexts and variables: they
are essentially the same as the ones we gave in \Cref{sec:intrinsictyping}.

Our type of term is indexed by a phase, a stage, a type
at that stage, and a context.

\ExecuteMetaData[BasicTwoTT.tex]{termdecl}

The first constructors are familiar: they are exactly the ones
seen in the previous section. These constructs are available
at both levels and both before and after staging hence the fact
that the phase and stage indices are polymorphic here.

\ExecuteMetaData[BasicTwoTT.tex]{termstlc}

Next we have the constructs specific to the two level calculus:
quotes (\AIC{`⟨\_⟩}) let users insert dynamic terms into static
expressions while splices (\AIC{`∼\_}) allow static terms to
be inserted in dynamic ones.
Staging will, by definition, eliminate these and so their phase
index is constrained to be \AIC{src}.

\ExecuteMetaData[BasicTwoTT.tex]{termtwolevel}

Putting it all together, we obtain the following inductive
family representing a minimal intrinsically typed two-level
calculus.

\begin{AgdaAlign}
  \begin{AgdaSuppressSpace}
    \ExecuteMetaData[BasicTwoTT.tex]{term}
  \end{AgdaSuppressSpace}
\end{AgdaAlign}

We can readily write examples such as the following definitions
of a purely dynamic and a purely static identity function. The
dynamic function will survive staging even if it is applied to a
dynamic argument while the static one can only exist in the
source phase and will be fully evaluated during staging.

\noindent
\begin{minipage}{.21\textwidth}
  \ExecuteMetaData[BasicTwoTT.tex]{iddyn}
\end{minipage}\hfill
\begin{minipage}{.23\textwidth}
  \ExecuteMetaData[BasicTwoTT.tex]{idsta}
\end{minipage}

Now that we have a syntax, we can start building the machinery
that will actually perform its partial evaluation.

\section{Staging by Evaluation}

The goal of this section is to define a type of \AF{Value}s
as well as an evaluation function which computes the
value associated to each term, provided that we have an
appropriate environment to interpret the term's free variables.
This will once again yield a function \AF{eval} of the following type.

\ExecuteMetaData[BasicTwoTT.tex]{evaldecl}

As a corollary we will obtain a staging function
that takes a closed dynamic term and gets rid of all of
the quotes and splices by fully
evaluating all of its static parts.

\ExecuteMetaData[BasicTwoTT.tex]{stagedecl}

We start with the model construction describing precisely
the type of values.

\subsection{Model Construction}\label{sec:stagingmodel}

The type of values is defined by case analysis on the stage.
Static values are given a static meaning (defined below)
while dynamic values are given a meaning as staged terms
i.e. terms guaranteed not to contain any static subterm.

\ExecuteMetaData[BasicTwoTT.tex]{model}

The family of static values is defined by induction on
the value's type. It is fairly similar to the standard
normalisation by evaluation construction
except that static values at a base types cannot possibly
be neutral terms.

\begin{AgdaSuppressSpace}
\ExecuteMetaData[BasicTwoTT.tex]{modelstadecl}
\ExecuteMetaData[BasicTwoTT.tex]{modelsta}
\end{AgdaSuppressSpace}

There are no static values of type \AIC{`α} as this base type does
not have any associated constructors and so we return the empty type \AD{⊥};
values of type (\AIC{`⇑} \AB{A}) are dynamic values of type \AB{A}
i.e. staged terms of type \AB{A};
functions from \AB{A} to \AB{B} are interpreted using Kripke function
spaces from static values of type \AB{A} to static values of type \AB{B}.

\subsection{Evaluation}

We can now explain what the meaning of each term constructor is.
In every instance we will proceed by case analysis on the
stage the meaning is being used at, essentially using a
meaning inspired by normalisation by evaluation for the static
part and one inspired by substitution for the dynamic one.

Application is interpreted as the semantic application defined
for Kripke function spaces in the static case, and the syntactic
\AIC{`app} constructor in the dynamic one.

\ExecuteMetaData[BasicTwoTT.tex]{app}

Lambda-abstractions are mapped to Kripke λs for static values
and to syntactic ones for the dynamic ones.

\ExecuteMetaData[BasicTwoTT.tex]{lam}

Putting it all together, we obtain the following definition
of the evaluation function.
Note that by virtue of the model construction the interpretation
of both \AIC{`∼\_} and \AIC{`⟨\_⟩} is the identity function:
static values of type (\AIC{`⇑} A) and staged terms of type \AB{A}
are interchangeable.

\begin{AgdaSuppressSpace}
\ExecuteMetaData[BasicTwoTT.tex]{evaldecl}
\ExecuteMetaData[BasicTwoTT.tex]{eval}
\end{AgdaSuppressSpace}

The function \AF{eval} is mutually defined with an auxiliary
function \AF{body} describing its behaviour on the body of a λ-abstraction.
It is defined using semantics lambdas and \AF{extend}.

\begin{AgdaSuppressSpace}
\ExecuteMetaData[BasicTwoTT.tex]{bodydecl}
\ExecuteMetaData[BasicTwoTT.tex]{body}
\end{AgdaSuppressSpace}

We finally obtain the \AF{stage} function by calling \AF{eval} with an
empty environment.

\begin{AgdaSuppressSpace}
\ExecuteMetaData[BasicTwoTT.tex]{stagefun}
\end{AgdaSuppressSpace}

\begin{remark}[(Co)Pattern-Matching Lambda]
The keyword (\AS{λ} \AK{where}) is analogous to Haskell's
\texttt{\textbackslash{}case}: it introduces a pattern-matching lambda.
In this instance, it is a copattern-matching one: we define
the environment of type (\AR{Env} \AIC{ε} \AIC{ε}) by
copattern-matching on {\AS{.}\ARF{get}} which allows us to bind
an argument of type (\AD{Var} \AB{A} \AIC{ε}) that can in turn
be immediately dismissed as uninhabited using the empty pattern \AS{()}.
\end{remark}

\section{A More Practical Two Level Calculus}

We are now going to extend the minimal calculus we used so far to
show a more realistic example of a two level calculus.

First we are going to add natural numbers and their eliminator.
These will be available at both stages and we will see how we
can transfer a static natural number to the dynamic phase by
defining a static \AF{`reify} term.

Second, based on Kov{\'{a}}cs' observation that the static and
dynamic language do not need to have exactly the same features,
we are going to add a type of static pairs.
These pairs and their projections can be used in arbitrary static
code but will be guaranteed to be evaluated away during staging.
We will demonstrate this by giving a static term \AF{`fib}
implementing a standard linear (ignoring the cost of addition)
algorithm for the Fibonacci function.
This will allow us to obtain e.g.

\ExecuteMetaData[TwoTT.tex]{testfib}

\noindent where \AF{fromℕ} is a helper function turning Agda
literals into \AF{Term}s built using \AIC{`zero} and \AIC{`succ},
and \AF{`add} is a dynamic addition function.
Note that the dynamic call to addition was not evaluated
away during staging.

\subsection{Adding Natural Numbers}

Our first extension adds the inductive type of Peano-style natural numbers,
its two constructors, and the appropriate eliminator for it.

\subsubsection{Types and Terms}

First we extend the definition of \AD{Type} with a new constructor
\AIC{`ℕ}. Natural numbers will be present at both stages and so we
allow the index to be polymorphic.

\ExecuteMetaData[TwoTT.tex]{typesnat}

We then add \AD{Term} constructors for the two Peano-style
constructors (\AIC{`zero} and \AIC{`succ}) as well as an
eliminator (\AIC{`iter}) which turns a natural number into
its Church encoding~\cite[Chapter 3]{church1941calculi}.

\ExecuteMetaData[TwoTT.tex]{termnat}

Our first program example is the function \AF{`reify} that
turns its static natural number argument into a dynamic
encoding. It does so by iterating over its input and replacing
static \AIC{`zero}s and \AIC{`succ}s by dynamic ones.

\ExecuteMetaData[TwoTT.tex]{reify}

We can also naturally define addition as iterated calls to
\AIC{`succ}. This definition is valid at both stages hence
the polymorphic phase and stage indices.

\label{def:add}
\ExecuteMetaData[TwoTT.tex]{add}

Let us now see how to evaluate the newly added constructs.

\subsubsection{Staging by Evaluation}

We extend the definition of \AF{Static} with a new clause decreeing
that values of type \AIC{`ℕ} are constant natural numbers.

\ExecuteMetaData[TwoTT.tex]{modelnat}

We can then describe the semantical counterparts of the newly
added constructors.
The term constructor \AIC{`zero} is either interpreted by
the natural number \AN{0} or by the term constructor itself
depending on whether it is used in a static or dynamic manner.

\ExecuteMetaData[TwoTT.tex]{zero}

Similarly \AIC{`succ} is interpreted either as (\AN{1} \AF{+}\AS{\_})
if it used in a static manner or by the term constructor itself
for dynamic uses.

\ExecuteMetaData[TwoTT.tex]{succ}

The meaning of \AIC{`iter} in the static layer is defined in
terms of the \AF{iterate} function defined by pattern-matching
in the host language and turning a natural number into its
Church encoding. Note that we need to use \AF{wkKripke} to
bring the \AB{succ} argument into the wider scope the \AB{zero}
one lives in.

\ExecuteMetaData[TwoTT.tex]{iter}

We can readily compute with these numbers. Reifying the static
result obtained by adding $7$ to $35$ will for instance return
$42$ (here \AF{fromℕ} once again stands for a helper function
turning Agda literals into \AD{Term} numbers).

\ExecuteMetaData[TwoTT.tex]{testadd}

Let us now look at an example of the fact, highlighted in
Kov{\'{a}}cs' original paper, that static datatypes do not
need to have a counterpart at runtime.

\subsection{Adding Static Pairs}\label{sec:stagingmodelprod}

We now want to add pairs that are only available in the static
layer and ensure that all traces of pairs and their projections
will have completely disappeared after staging.

\subsubsection{Types and Terms}

We first extend the inductive definition of object types
with a new construct for pair types. It is explicitly
marked as static (\AIC{sta}) only.

\ExecuteMetaData[TwoTT.tex]{typesprod}

We then extend the inductive family of term constructs
with a constructor for pairs (\AIC{\_`,\_}) and two constructors
for the first (\AIC{`fst}) and second (\AIC{`snd}) projection
respectively.

\ExecuteMetaData[TwoTT.tex]{termprod}

This enables us to implement in the static layer the classic
linear definition of the Fibonacci function which internally
uses a pair of the current Fibonacci number and its successor.
It is obtained by taking the first projection of the result
of iterating the invariant-respecting \emph{step} function
over the valid \emph{base} case.

\ExecuteMetaData[TwoTT.tex]{fib}

This definition uses \AF{\_`∘\_} defined in~\Cref{def:composition},
and \AF{`add} defined in~\Cref{def:add}.

\subsubsection{Staging by Evaluation}

The amendment to the model construction and the definition of the
constructors' semantical counterparts is easy.
First, static pairs are pairs of static values.

\ExecuteMetaData[TwoTT.tex]{modelprod}

Second, pair constructors are mapped to pair constructors in
the host language, and the same for projections.

\ExecuteMetaData[TwoTT.tex]{evalprod}

These definition now allow us to evaluate static calls to the Fibonacci
function such as the one presented in this section's introduction.

\ExecuteMetaData[TwoTT.tex]{testfib}

While this addition of static pairs may seem interesting
but anecdotal, the same techniques can be used to work on
defining a much more applicable two level language.

\section{Application: Circuit Generation}\label{sec:circuits}

This section's content is inspired by Quipper, a functional
programming language to describe quantum computations
introduced by Green, Lumsdaine, Ross, Selinger, and
Valiron~\cite{DBLP:conf/rc/GreenLRSV13} and related
formal treatments such as Rennela and Staton's categorical
models~\cite{DBLP:journals/lmcs/RennelaS19}.
This strand of research gives us a good example of a setting in which
we have two very distinct layers: a static layer with a
full-fledged functional language, and a dynamic layer of
quantum circuits obtained by partially evaluating the source.

In our proof of concept, we study a minimal language of
classical circuits inspired by Π-ware a formal hardware
description and verification language proposed by
Flor, Swierstra, and Sijsling~\cite{DBLP:conf/types/FlorSS15}.
This allows us to focus on the two-level aspect instead of
having to deal with linearity and unitary operators which are
specific to the Quantum setting.

\subsection{Types and Terms}

Our definition of types should now be mostly unsurprising.
We have function spaces (this time confined to the static
layer), a lifting construct allowing the embedding of
dynamic types in the static layer at the source stage,
and finally a type of circuits
\AIC{`⟨}~\AB{i}~\AIC{∣}~\AB{o}~\AIC{⟩} characterised by
their input (\AB{i}) and output (\AB{o}) arities,
each represented by a natural number in the host language.

\ExecuteMetaData[MetaCircuit.tex]{type}

Next, we extend the basic simply typed lambda calculus with
quotes and splices with term constructors for circuit descriptions.
They will all belong to the dynamic stage.
Our first constructor gives us the universal nand gate.
Its type records the fact it takes two inputs and returns
a single output.

\ExecuteMetaData[MetaCircuit.tex]{termcircuitnand}

Next, we have a constructor for the parallel composition
of existing circuits. The input and output arities of the
resulting circuit are obtained by adding up the respective
input and output arities of each of the components.

\ExecuteMetaData[MetaCircuit.tex]{termcircuitpar}

We can also compose circuits sequentially, provided
that the output arity of the first circuit matches
the input arity of the second.

\ExecuteMetaData[MetaCircuit.tex]{termcircuitseq}

Finally, we follow the Π-ware~\cite{DBLP:conf/types/FlorSS15}
approach and offer a general rewiring component.
A `mix' of $i$ inputs returning $o$ outputs is defined by
a vector (i.e. a list of known length) of size $o$
containing finite numbers between $0$ and $i$
corresponding to the input the output is connected to.
This allows arbitrary duplications and deletions
of inputs.

\ExecuteMetaData[MetaCircuit.tex]{termcircuitmix}

Typical examples include
\AF{`id$_2$} (the \emph{id}entity circuit on two inputs),
\AF{`swap} (the circuit \emph{swap}ping its two inputs),
and \AF{`dup} (the circuit \emph{dup}licating its single input).
We present them below together with the corresponding wiring diagrams.

\noindent
\begin{minipage}{.25\textwidth}
  \ExecuteMetaData[MetaCircuit.tex]{id2}
\end{minipage}\hfill
\begin{minipage}{.125\textwidth}
  \begin{tikzpicture}
    \draw[fill] (0,0)  circle[radius=1.5pt] node { };
    \draw[fill] (0,.5) circle[radius=1.5pt] node { };

    \draw[fill] (1,0)  circle[radius=1.5pt] node { };
    \draw[fill] (1,.5) circle[radius=1.5pt] node { };

    \draw[-] (0,0)  to [out=0,in=180] (.95,0);
    \draw[-] (0,.5) to [out=0,in=180] (.95,.5);
  \end{tikzpicture}
\end{minipage}

\noindent
\begin{minipage}{.25\textwidth}
  \ExecuteMetaData[MetaCircuit.tex]{swap}
\end{minipage}\hfill
\begin{minipage}{.125\textwidth}
  \begin{tikzpicture}
    \draw[fill] (0,0)  circle[radius=1.5pt] node { };
    \draw[fill] (0,.5) circle[radius=1.5pt] node { };

    \draw[fill] (1,0)  circle[radius=1.5pt] node { };
    \draw[fill] (1,.5) circle[radius=1.5pt] node { };

    \draw[-] (0,0)  to [out=0,in=180] (.95,.5);
    \draw[-] (0,.5) to [out=0,in=180] (.95,0);
  \end{tikzpicture}
\end{minipage}

\noindent
\begin{minipage}{.25\textwidth}
  \ExecuteMetaData[MetaCircuit.tex]{dup}
\end{minipage}\hfill
\begin{minipage}{.125\textwidth}
  \begin{tikzpicture}
    \draw[fill] (0,.25) circle[radius=1.5pt] node { };

    \draw[fill] (1,0)  circle[radius=1.5pt] node { };
    \draw[fill] (1,.5) circle[radius=1.5pt] node { };

    \draw[-] (0,.25) to [out=-45,in=180] (.95,0);
    \draw[-] (0,.25) to [out=45,in=180] (.95,.5);
  \end{tikzpicture}
\end{minipage}

We can then define our first real example: \AF{`diag},
a static program taking a circuit with two inputs and
one output and returning a circuit with one input and
one output.
It does so by first duplicating the one input using \AF{`dup}
and then feeding it to both of the argument's ports.
We present it below together with the corresponding
circuit diagram.

\ExecuteMetaData[MetaCircuit.tex]{diag}

\medskip
\begin{minipage}{.05\textwidth}
  $c \mapsto$
\end{minipage}
\begin{minipage}{.375\textwidth}
  \begin{tikzpicture}

    \coordinate (x) at (0,.25);
    \draw[fill] (x) circle[radius=1.5pt] node {};
    \draw (x)+(-.5,0) node {$x$} edge (x);

    \node[rectangle, draw, minimum height=.75cm, minimum width=.5cm] at (1, .25) (c) {$c$};

    \draw[-] (x) to [out=-45,in=180] ([yshift=.1cm] c.south west);
    \draw[-] (x) to [out=45,in=180] ([yshift=-.1cm] c.north west);

    \coordinate (r) at ([xshift=1cm] c);
    \draw[fill] (r) circle[radius=1.5pt] node {};
    \draw[-] (c.east) to (r);
    \draw (r)+(.5,0) node {$r$} edge (r);
  \end{tikzpicture}
\end{minipage}
\medskip

We can then obtain the \AF{`not} gate by taking the diagonal
of the \AIC{`nand} built-in gate.

\ExecuteMetaData[MetaCircuit.tex]{not}

Staging this definition does evaluate away all of the function
calls to yield a simple circuit obtained by sequentially
composing \AF{`dup} and \AIC{`nand} as shown below.

\ExecuteMetaData[MetaCircuit.tex]{testNot}

Using standard constructions, we can define \AF{`and} and \AF{`or}
in terms of the universal \AIC{`nand} gate.

\noindent
\begin{minipage}{.22\textwidth}
  \ExecuteMetaData[MetaCircuit.tex]{and}
\end{minipage}\hfill
\begin{minipage}{.22\textwidth}
  \ExecuteMetaData[MetaCircuit.tex]{or}
\end{minipage}

Going back to a slightly more complex setting, adding booleans in the
static layer lets us once again define more interesting terms.
For instance, the following \AF{`tab} circuit \emph{tab}ulating its
input: given a function that takes a boolean and computes a one-input
one-output circuit, it returns a circuit with two inputs and one
output that has the same behaviour.

\ExecuteMetaData[MetaCircuit.tex]{tab}

Using dashed lines to separate the different constituting parts
of the circuit as defined above, we obtain the following circuit
diagram.

\medskip
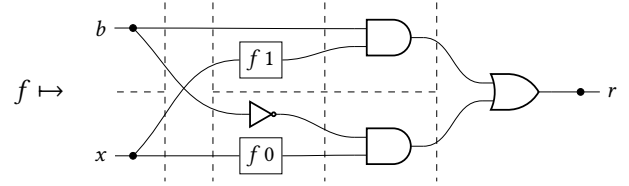
\begin{minipage}{.05\textwidth}
  $f \mapsto$
\end{minipage}\hfill
\begin{minipage}{.425\textwidth}\scalebox{.85}{
  \begin{tikzpicture}{circuit logic US}
    \coordinate (b) at (0,2);
    \draw[fill] (b) circle[radius=1.5pt] node {};
    \draw (b)+(-.5,0) node {$b$} edge (b);

    \coordinate (x) at (0,0);
    \draw[fill] (x) circle[radius=1.5pt] node {};
    \draw (x)+(-.5,0) node {$x$} edge (x);

    \node[rectangle, draw] at (2, 1.5) (f1) {$f\,1$};
    \draw[-] (x) to [out=45, in=180] (f1);

    \node[rectangle, draw] at (2, 0) (f0) {$f\,0$};
    \draw[-] (x) to [out=0, in=180] (f0);

    \draw (4, 0.15) node[and port, scale=.5] (and0) {};
    \draw[-] (f0) [out=0, in=180] to (and0.in 2);

    \draw (4, 1.85) node[and port, scale=.5] (and1) {};
    \draw[-] (b) [out=0, in=180] to (and1.in 1);
    \draw[-] (f1) [out=0, in=180] to (and1.in 2);

    \draw (2, .65) node[not port, scale=.35] (not0) {};
    \draw[-] (b) [out=-45, in=180] to (not0.in 1);
    \draw[-] (not0.out) [out=0, in=180] to (and0.in 1);

    \draw (6, 1) node[or port, scale=.5] (or01) {};
    \draw[-] (and1.out) [out=0, in=180] to (or01.in 1);
    \draw[-] (and0.out) [out=0, in=180] to (or01.in 2);

    \coordinate (tag) at (7,1);
    \draw[fill] (tag) circle[radius=1.5pt] node {};
    \draw (tag)+(.5,0) node {$r$} edge (tag);
    \draw[-] (or01.out) to (tag);

    \draw[dashed] ([xshift=-.25cm, yshift=1cm] x) to ([xshift=.5cm, yshift=1cm] x);
    \draw[dashed] ([xshift=.5cm, yshift=.4cm] b) to ([xshift=.5cm, yshift=-.4cm] x);
    \draw[dashed] ([xshift=1.25cm, yshift=.4cm] b) to ([xshift=1.25cm, yshift=-.4cm] x);
    \draw[dashed] ([xshift=3cm, yshift=.4cm] b) to ([xshift=3cm, yshift=-.4cm] x);
    \draw[dashed] ([xshift=4.75cm, yshift=.4cm] b) to ([xshift=4.75cm, yshift=-.4cm] x);
    \draw[dashed] ([xshift=1.25cm, yshift=1cm] x) to ([xshift=4.75cm, yshift=1cm] x);

  \end{tikzpicture}}
\end{minipage}
\medskip

This term is not in and of itself particularly useful but its
generalisation to one that could take a function computing an
{\AIC{⟨}~\AB{i}~\AIC{∣}~\AB{o}~\AIC{⟩}} circuit and return an
equivalent {\AIC{⟨}~\AN{1}~\AF{+}~\AB{i}~\AIC{∣}~\AB{o}~\AIC{⟩}}
circuit would allow us to build arbitrarily complex circuits
by tabulating static n-ary boolean functions.

This would however require a setting where the static layer is
dependently typed like in Kov{\'{a}}cs' original work, something
out of scope for this paper.

\section{Related Work}

Prior work on partial evaluation and metaprogramming
abounds so we will only focus on the very most relevant
works involving strong types.

\subsection{Partial Evaluation as Code Generation}

Using partial evaluation as a code generation strategy
is not new. Despite not being stated in two-level terms,
Scholz~\cite{DBLP:conf/pepm/Scholz14} for instance
describes the extensive use of partial evaluation backed
by a multi-level type system in a single assignment C
compiler as a way to generate efficient code.
Instead of relying on expensive static analysis (of e.g.
the array shapes that are statically known) to generate
highly specialised code, they produce the most generic
code and then use a powerful type-directed partial
evaluation pass to residualise it.
This has the additional benefit of subsuming other
compiler passes such as constant folding.

\subsection{Typed Metaprogramming}

Carette and Kiselyov~\cite{DBLP:journals/scp/CaretteK11}
use the many Gaussian Elimination variants as a case study
in hygienic, well typed, modular, and performant multi-stage
programming in MetaOCaml.
They underline that MetaOCaml is purely generative i.e.
that the staged code cannot be inspected during staging
itself.
In contrast, and as we explain in the future works
(\Cref{sec:computrel}), the two-level setup is not
intrinsically tied to computationally irrelevant quotes
and splices.

Jang, Gélineau, Monnier, and Pientka's
Mœbius~\cite{DBLP:journals/pacmpl/JangGMP22} defines
a type theory with a built in notion of quasiquotations
that can be used to generate programs in a type-safe
manner.
Unlike MetaOCaml, this language lets metaprograms
inspect the code fragments
they are passed as arguments thus allowing e.g. the
implementation of optimisation passes post-processing the
result of a prior metaprogram.
This is extremely powerful, at the cost of a more complex
underlying theory.
In Mœbius the meta and object language are essentially the
same but it does not seem to be a necessary restriction.

\subsection{Quantum Circuits Generation}
As already mentioned in \Cref{sec:circuits}, such two level
systems occur naturally when defining high level languages
for (quantum) circuit descriptions.
Rennela and Staton's EWire language is itself the categorical
treatment of a minor generalisation of Paykin, Rand,
and Zdancewic's QWire~\cite{DBLP:conf/popl/Paykin0Z17},
a clear invariant-enforcing improvement over the weakly
typed Haskell embedded domain specific language Quipper~\cite{DBLP:conf/rc/GreenLRSV13}.
EWire is an ad-hoc construction which, although not worded
explicitly in terms of a two level type theory, effectively
is one: quotes and splices are called boxing and unboxing,
and a QWire-inspired partial normalisation procedure
proven to be semantics-preserving is defined.
The formal well scoped-and-typed treatment of these languages
can be built on top of Wood and Atkey's work on linear
meta-theory~\cite{DBLP:conf/esop/WoodA22}.

\subsection{SMT Constraints Generation}
In their work on compiling higher order specifications to SMT
constraints~\cite{DBLP:conf/cpp/DaggittAKKA23},
Daggit, Atkey, Kokke, Komendantskaya, and Arnaboldi
designed a cunning 'translation by evaluation' to
partially evaluate specifications written in a full featured
high level functional language (without recursion)
into first order SMT constraints.
This is not explicitly designed as a two level system and
so the sucess of the partial evaluation comes from a careful
but ultimately ad-hoc design rather than a systematic approach.
Unlike ours, their system however comes with a proof of
correctness: the generated formula is proven to be logically
equivalent to the high level specification.
This is an obvious avenue for future work on our part.

\section{Future Work}

\subsection{Soundness and Completeness}
We focused here on the intrinsically typed language description,
the corresponding model construction, and the acquisition of a
staging-by-evaluation function as a corollary.
Following Catarina Coquand's work on formalising normalisation
by evaluation~\cite{DBLP:journals/lisp/Coquand02} we could
additionally introduce the appropriate logical relations to
prove that this process is sound and complete with respect
to a small step semantics for the static layer.

\subsection{Computationally Relevant Quotes and Splices}\label{sec:computrel}
In all of the examples we have seen, the interpretation of
quotes and splices in the model is the identity function.
This is not however a necessary constraint and we could
imagine definitions of analogues to quotes and splices
with interesting metaprogramming uses.
We have for instance a proof of concept extending our circuit
language with vectors of booleans (\AIC{`[} \AB{i} \AIC{]} stands
for a vector of size \AB{i}) and two additional primitives
\AIC{`run} and \AIC{`tab} converting between circuits and static
functions over vectors.
They have the following types and are interpreted in the model
by symbolically evaluating the circuit for \AIC{`run}, and
tabulating the function for \AIC{`tab}.

\ExecuteMetaData[RunMetaCircuit.tex]{runtab}

The circuits generated by partial evaluation are currently
unoptimised but this demonstrates the feasibility of adding
constructs mediating between the type theory's two levels
which have a non-trivial semantics.
We want to identify more opportunities for such computationally
relevant analogues to quotes and splices.

\subsection{Dependently Typed Circuit Description Language}
Our undergraduates are already being taught digital logic
using a functional-style circuit description language.
Extending it with a dependently typed meta-programming
layer would allow them to structure their understanding
of the generic construction of arithmetic circuits for
arbitrarily large inputs.

\subsection{Generic Two Level Construction}
Even though we have seen that having two wildly different
language layers can be extremely useful, a two-level
construction with exactly the same features is still very
interesting: it lets programmers use their language of choice
as its own metaprogramming facilities.
Correspondingly, giving a generic treatment of the construction
taking a language and returning its standard two-level version
is an important endeavour.
A promising approach involves defining such a transformation
by induction over a universe of language
descriptions~\cite{DBLP:journals/jfp/AllaisACMM21}.

%%
%% The acknowledgments section is defined using the "acks" environment
%% (and NOT an unnumbered section). This ensures the proper
%% identification of the section in the article metadata, and the
%% consistent spelling of the heading.
\begin{acks}
\iftoggle{BLIND}{
  Thanks to XXX for comments.
}{
  We would like to thank Bob Atkey for his suggestion to index
  stages by a phase, thus allowing us to ensure that a staged
  term does not have any static subterm.

  We would additionally like to thank the reviewers for their
  helpful suggestions of relevant related works.
}

\end{acks}

\balance
%%
%% The next two lines define the bibliography style to be used, and
%% the bibliography file.
\bibliographystyle{ACM-Reference-Format}
\bibliography{staging}

%%% -*-BibTeX-*-
%%% Do NOT edit. File created by BibTeX with style
%%% ACM-Reference-Format-Journals [18-Jan-2012].

\begin{thebibliography}{26}

%%% ====================================================================
%%% NOTE TO THE USER: you can override these defaults by providing
%%% customized versions of any of these macros before the \bibliography
%%% command.  Each of them MUST provide its own final punctuation,
%%% except for \shownote{}, \showDOI{}, and \showURL{}.  The latter two
%%% do not use final punctuation, in order to avoid confusing it with
%%% the Web address.
%%%
%%% To suppress output of a particular field, define its macro to expand
%%% to an empty string, or better, \unskip, like this:
%%%
%%% \newcommand{\showDOI}[1]{\unskip}   % LaTeX syntax
%%%
%%% \def \showDOI #1{\unskip}           % plain TeX syntax
%%%
%%% ====================================================================

\ifx \showCODEN    \undefined \def \showCODEN     #1{\unskip}     \fi
\ifx \showDOI      \undefined \def \showDOI       #1{#1}\fi
\ifx \showISBNx    \undefined \def \showISBNx     #1{\unskip}     \fi
\ifx \showISBNxiii \undefined \def \showISBNxiii  #1{\unskip}     \fi
\ifx \showISSN     \undefined \def \showISSN      #1{\unskip}     \fi
\ifx \showLCCN     \undefined \def \showLCCN      #1{\unskip}     \fi
\ifx \shownote     \undefined \def \shownote      #1{#1}          \fi
\ifx \showarticletitle \undefined \def \showarticletitle #1{#1}   \fi
\ifx \showURL      \undefined \def \showURL       {\relax}        \fi
% The following commands are used for tagged output and should be
% invisible to TeX
\providecommand\bibfield[2]{#2}
\providecommand\bibinfo[2]{#2}
\providecommand\natexlab[1]{#1}
\providecommand\showeprint[2][]{arXiv:#2}

\bibitem[Abel et~al\mbox{.}(2013)]%
        {DBLP:conf/popl/AbelPTS13}
\bibfield{author}{\bibinfo{person}{Andreas Abel}, \bibinfo{person}{Brigitte
  Pientka}, \bibinfo{person}{David Thibodeau}, {and} \bibinfo{person}{Anton
  Setzer}.} \bibinfo{year}{2013}\natexlab{}.
\newblock \showarticletitle{Copatterns: programming infinite structures by
  observations}. In \bibinfo{booktitle}{\emph{The 40th Annual {ACM}
  {SIGPLAN-SIGACT} Symposium on Principles of Programming Languages, {POPL}
  '13, Rome, Italy - January 23 - 25, 2013}},
  \bibfield{editor}{\bibinfo{person}{Roberto Giacobazzi} {and}
  \bibinfo{person}{Radhia Cousot}} (Eds.). \bibinfo{publisher}{{ACM}},
  \bibinfo{pages}{27--38}.
\newblock
\urldef\tempurl%
\url{https://doi.org/10.1145/2429069.2429075}
\showDOI{\tempurl}


\bibitem[Allais et~al\mbox{.}(2021)]%
        {DBLP:journals/jfp/AllaisACMM21}
\bibfield{author}{\bibinfo{person}{Guillaume Allais}, \bibinfo{person}{Robert
  Atkey}, \bibinfo{person}{James Chapman}, \bibinfo{person}{Conor McBride},
  {and} \bibinfo{person}{James McKinna}.} \bibinfo{year}{2021}\natexlab{}.
\newblock \showarticletitle{A type- and scope-safe universe of syntaxes with
  binding: their semantics and proofs}.
\newblock \bibinfo{journal}{\emph{J. Funct. Program.}}  \bibinfo{volume}{31}
  (\bibinfo{year}{2021}), \bibinfo{pages}{e22}.
\newblock
\urldef\tempurl%
\url{https://doi.org/10.1017/S0956796820000076}
\showDOI{\tempurl}


\bibitem[Allais et~al\mbox{.}(2017)]%
        {DBLP:conf/cpp/Allais0MM17}
\bibfield{author}{\bibinfo{person}{Guillaume Allais}, \bibinfo{person}{James
  Chapman}, \bibinfo{person}{Conor McBride}, {and} \bibinfo{person}{James
  McKinna}.} \bibinfo{year}{2017}\natexlab{}.
\newblock \showarticletitle{Type-and-scope safe programs and their proofs}. In
  \bibinfo{booktitle}{\emph{Proceedings of the 6th {ACM} {SIGPLAN} Conference
  on Certified Programs and Proofs, {CPP} 2017, Paris, France, January 16-17,
  2017}}, \bibfield{editor}{\bibinfo{person}{Yves Bertot} {and}
  \bibinfo{person}{Viktor Vafeiadis}} (Eds.). \bibinfo{publisher}{{ACM}},
  \bibinfo{pages}{195--207}.
\newblock
\urldef\tempurl%
\url{https://doi.org/10.1145/3018610.3018613}
\showDOI{\tempurl}


\bibitem[Altenkirch et~al\mbox{.}(1995)]%
        {DBLP:conf/ctcs/AltenkirchHS95}
\bibfield{author}{\bibinfo{person}{Thorsten Altenkirch},
  \bibinfo{person}{Martin Hofmann}, {and} \bibinfo{person}{Thomas Streicher}.}
  \bibinfo{year}{1995}\natexlab{}.
\newblock \showarticletitle{Categorical Reconstruction of a Reduction Free
  Normalization Proof}. In \bibinfo{booktitle}{\emph{Category Theory and
  Computer Science, 6th International Conference, {CTCS} '95, Cambridge, UK,
  August 7-11, 1995, Proceedings}} \emph{(\bibinfo{series}{Lecture Notes in
  Computer Science}, Vol.~\bibinfo{volume}{953})},
  \bibfield{editor}{\bibinfo{person}{David~H. Pitt}, \bibinfo{person}{David~E.
  Rydeheard}, {and} \bibinfo{person}{Peter~T. Johnstone}} (Eds.).
  \bibinfo{publisher}{Springer}, \bibinfo{pages}{182--199}.
\newblock
\urldef\tempurl%
\url{https://doi.org/10.1007/3-540-60164-3_27}
\showDOI{\tempurl}


\bibitem[Altenkirch and Reus(1999)]%
        {DBLP:conf/csl/AltenkirchR99}
\bibfield{author}{\bibinfo{person}{Thorsten Altenkirch} {and}
  \bibinfo{person}{Bernhard Reus}.} \bibinfo{year}{1999}\natexlab{}.
\newblock \showarticletitle{Monadic Presentations of Lambda Terms Using
  Generalized Inductive Types}. In \bibinfo{booktitle}{\emph{Computer Science
  Logic, 13th International Workshop, {CSL} '99, 8th Annual Conference of the
  EACSL, Madrid, Spain, September 20-25, 1999, Proceedings}}
  \emph{(\bibinfo{series}{Lecture Notes in Computer Science},
  Vol.~\bibinfo{volume}{1683})}, \bibfield{editor}{\bibinfo{person}{J{\"{o}}rg
  Flum} {and} \bibinfo{person}{Mario Rodr{{\i}}guez{-}Artalejo}} (Eds.).
  \bibinfo{publisher}{Springer}, \bibinfo{pages}{453--468}.
\newblock
\urldef\tempurl%
\url{https://doi.org/10.1007/3-540-48168-0\_32}
\showDOI{\tempurl}


\bibitem[Bellegarde and Hook(1994)]%
        {DBLP:journals/scp/BellegardeH94}
\bibfield{author}{\bibinfo{person}{Fran{\c{c}}oise Bellegarde} {and}
  \bibinfo{person}{James Hook}.} \bibinfo{year}{1994}\natexlab{}.
\newblock \showarticletitle{Substitution: {A} Formal Methods Case Study Using
  Monads and Transformations}.
\newblock \bibinfo{journal}{\emph{Sci. Comput. Program.}} \bibinfo{volume}{23},
  \bibinfo{number}{2-3} (\bibinfo{year}{1994}), \bibinfo{pages}{287--311}.
\newblock
\urldef\tempurl%
\url{https://doi.org/10.1016/0167-6423(94)00022-0}
\showDOI{\tempurl}


\bibitem[Berger and Schwichtenberg(1991)]%
        {DBLP:conf/lics/BergerS91}
\bibfield{author}{\bibinfo{person}{Ulrich Berger} {and} \bibinfo{person}{Helmut
  Schwichtenberg}.} \bibinfo{year}{1991}\natexlab{}.
\newblock \showarticletitle{An Inverse of the Evaluation Functional for Typed
  lambda-calculus}. In \bibinfo{booktitle}{\emph{Proceedings of the Sixth
  Annual Symposium on Logic in Computer Science {(LICS} '91), Amsterdam, The
  Netherlands, July 15-18, 1991}}. \bibinfo{publisher}{{IEEE} Computer
  Society}, \bibinfo{pages}{203--211}.
\newblock
\urldef\tempurl%
\url{https://doi.org/10.1109/LICS.1991.151645}
\showDOI{\tempurl}


\bibitem[Bird and Paterson(1999)]%
        {DBLP:journals/jfp/BirdP99}
\bibfield{author}{\bibinfo{person}{Richard~S. Bird} {and} \bibinfo{person}{Ross
  Paterson}.} \bibinfo{year}{1999}\natexlab{}.
\newblock \showarticletitle{De Bruijn Notation as a Nested Datatype}.
\newblock \bibinfo{journal}{\emph{J. Funct. Program.}} \bibinfo{volume}{9},
  \bibinfo{number}{1} (\bibinfo{year}{1999}), \bibinfo{pages}{77--91}.
\newblock
\urldef\tempurl%
\url{https://doi.org/10.1017/s0956796899003366}
\showDOI{\tempurl}


\bibitem[Carette and Kiselyov(2011)]%
        {DBLP:journals/scp/CaretteK11}
\bibfield{author}{\bibinfo{person}{Jacques Carette} {and} \bibinfo{person}{Oleg
  Kiselyov}.} \bibinfo{year}{2011}\natexlab{}.
\newblock \showarticletitle{Multi-stage programming with functors and monads:
  Eliminating abstraction overhead from generic code}.
\newblock \bibinfo{journal}{\emph{Sci. Comput. Program.}} \bibinfo{volume}{76},
  \bibinfo{number}{5} (\bibinfo{year}{2011}), \bibinfo{pages}{349--375}.
\newblock
\urldef\tempurl%
\url{https://doi.org/10.1016/J.SCICO.2008.09.008}
\showDOI{\tempurl}


\bibitem[Church(1941)]%
        {church1941calculi}
\bibfield{author}{\bibinfo{person}{Alonzo Church}.}
  \bibinfo{year}{1941}\natexlab{}.
\newblock \bibinfo{booktitle}{\emph{The calculi of lambda-conversion}}.
\newblock Number~6 in \bibinfo{series}{Annals of Mathematics Studies}.
  \bibinfo{publisher}{Princeton University Press}.
\newblock


\bibitem[Coquand(2002)]%
        {DBLP:journals/lisp/Coquand02}
\bibfield{author}{\bibinfo{person}{Catarina Coquand}.}
  \bibinfo{year}{2002}\natexlab{}.
\newblock \showarticletitle{A Formalised Proof of the Soundness and
  Completeness of a Simply Typed Lambda-Calculus with Explicit Substitutions}.
\newblock \bibinfo{journal}{\emph{High. Order Symb. Comput.}}
  \bibinfo{volume}{15}, \bibinfo{number}{1} (\bibinfo{year}{2002}),
  \bibinfo{pages}{57--90}.
\newblock
\urldef\tempurl%
\url{https://doi.org/10.1023/A:1019964114625}
\showDOI{\tempurl}


\bibitem[Coquand and Dybjer(1997)]%
        {DBLP:journals/mscs/CoquandD97}
\bibfield{author}{\bibinfo{person}{Thierry Coquand} {and}
  \bibinfo{person}{Peter Dybjer}.} \bibinfo{year}{1997}\natexlab{}.
\newblock \showarticletitle{Intuitionistic Model Constructions and
  Normalization Proofs}.
\newblock \bibinfo{journal}{\emph{Math. Struct. Comput. Sci.}}
  \bibinfo{volume}{7}, \bibinfo{number}{1} (\bibinfo{year}{1997}),
  \bibinfo{pages}{75--94}.
\newblock
\urldef\tempurl%
\url{https://doi.org/10.1017/S0960129596002150}
\showDOI{\tempurl}


\bibitem[Daggitt et~al\mbox{.}(2023)]%
        {DBLP:conf/cpp/DaggittAKKA23}
\bibfield{author}{\bibinfo{person}{Matthew~L. Daggitt}, \bibinfo{person}{Robert
  Atkey}, \bibinfo{person}{Wen Kokke}, \bibinfo{person}{Ekaterina
  Komendantskaya}, {and} \bibinfo{person}{Luca Arnaboldi}.}
  \bibinfo{year}{2023}\natexlab{}.
\newblock \showarticletitle{Compiling Higher-Order Specifications to {SMT}
  Solvers: How to Deal with Rejection Constructively}. In
  \bibinfo{booktitle}{\emph{Proceedings of the 12th {ACM} {SIGPLAN}
  International Conference on Certified Programs and Proofs, {CPP} 2023,
  Boston, MA, USA, January 16-17, 2023}},
  \bibfield{editor}{\bibinfo{person}{Robbert Krebbers},
  \bibinfo{person}{Dmitriy Traytel}, \bibinfo{person}{Brigitte Pientka}, {and}
  \bibinfo{person}{Steve Zdancewic}} (Eds.). \bibinfo{publisher}{{ACM}},
  \bibinfo{pages}{102--120}.
\newblock
\urldef\tempurl%
\url{https://doi.org/10.1145/3573105.3575674}
\showDOI{\tempurl}


\bibitem[de~Bruijn(1972)]%
        {de1972lambda}
\bibfield{author}{\bibinfo{person}{Nicolaas~Govert de Bruijn}.}
  \bibinfo{year}{1972}\natexlab{}.
\newblock \showarticletitle{Lambda {C}alculus Notation with Nameless Dummies}.
  In \bibinfo{booktitle}{\emph{Indagationes Mathematicae}},
  Vol.~\bibinfo{volume}{75}. Elsevier, \bibinfo{pages}{381--392}.
\newblock


\bibitem[Dybjer(1994)]%
        {DBLP:journals/fac/Dybjer94}
\bibfield{author}{\bibinfo{person}{Peter Dybjer}.}
  \bibinfo{year}{1994}\natexlab{}.
\newblock \showarticletitle{Inductive Families}.
\newblock \bibinfo{journal}{\emph{Formal Aspects Comput.}} \bibinfo{volume}{6},
  \bibinfo{number}{4} (\bibinfo{year}{1994}), \bibinfo{pages}{440--465}.
\newblock
\urldef\tempurl%
\url{https://doi.org/10.1007/BF01211308}
\showDOI{\tempurl}


\bibitem[Flor et~al\mbox{.}(2015)]%
        {DBLP:conf/types/FlorSS15}
\bibfield{author}{\bibinfo{person}{Jo{\~{a}}o Paulo~Pizani Flor},
  \bibinfo{person}{Wouter Swierstra}, {and} \bibinfo{person}{Yorick Sijsling}.}
  \bibinfo{year}{2015}\natexlab{}.
\newblock \showarticletitle{Pi-Ware: Hardware Description and Verification in
  Agda}. In \bibinfo{booktitle}{\emph{21st International Conference on Types
  for Proofs and Programs, {TYPES} 2015, May 18-21, 2015, Tallinn, Estonia}}
  \emph{(\bibinfo{series}{LIPIcs}, Vol.~\bibinfo{volume}{69})},
  \bibfield{editor}{\bibinfo{person}{Tarmo Uustalu}} (Ed.).
  \bibinfo{publisher}{Schloss Dagstuhl - Leibniz-Zentrum f{\"{u}}r Informatik},
  \bibinfo{pages}{9:1--9:27}.
\newblock
\urldef\tempurl%
\url{https://doi.org/10.4230/LIPIcs.TYPES.2015.9}
\showDOI{\tempurl}


\bibitem[Green et~al\mbox{.}(2013)]%
        {DBLP:conf/rc/GreenLRSV13}
\bibfield{author}{\bibinfo{person}{Alexander~S. Green},
  \bibinfo{person}{Peter~LeFanu Lumsdaine}, \bibinfo{person}{Neil~J. Ross},
  \bibinfo{person}{Peter Selinger}, {and} \bibinfo{person}{Beno{\^{\i}}t
  Valiron}.} \bibinfo{year}{2013}\natexlab{}.
\newblock \showarticletitle{An Introduction to Quantum Programming in Quipper}.
  In \bibinfo{booktitle}{\emph{Reversible Computation - 5th International
  Conference, {RC} 2013, Victoria, BC, Canada, July 4-5, 2013. Proceedings}}
  \emph{(\bibinfo{series}{Lecture Notes in Computer Science},
  Vol.~\bibinfo{volume}{7948})}, \bibfield{editor}{\bibinfo{person}{Gerhard~W.
  Dueck} {and} \bibinfo{person}{D.~Michael Miller}} (Eds.).
  \bibinfo{publisher}{Springer}, \bibinfo{pages}{110--124}.
\newblock
\urldef\tempurl%
\url{https://doi.org/10.1007/978-3-642-38986-3_10}
\showDOI{\tempurl}


\bibitem[Jang et~al\mbox{.}(2022)]%
        {DBLP:journals/pacmpl/JangGMP22}
\bibfield{author}{\bibinfo{person}{Junyoung Jang}, \bibinfo{person}{Samuel
  G{\'{e}}lineau}, \bibinfo{person}{Stefan Monnier}, {and}
  \bibinfo{person}{Brigitte Pientka}.} \bibinfo{year}{2022}\natexlab{}.
\newblock \showarticletitle{M{\oe}bius: metaprogramming using contextual types:
  the stage where system f can pattern match on itself}.
\newblock \bibinfo{journal}{\emph{Proc. {ACM} Program. Lang.}}
  \bibinfo{volume}{6}, \bibinfo{number}{{POPL}} (\bibinfo{year}{2022}),
  \bibinfo{pages}{1--27}.
\newblock
\urldef\tempurl%
\url{https://doi.org/10.1145/3498700}
\showDOI{\tempurl}


\bibitem[Kov{\'{a}}cs(2022)]%
        {DBLP:journals/pacmpl/Kovacs22}
\bibfield{author}{\bibinfo{person}{Andr{\'{a}}s Kov{\'{a}}cs}.}
  \bibinfo{year}{2022}\natexlab{}.
\newblock \showarticletitle{Staged compilation with two-level type theory}.
\newblock \bibinfo{journal}{\emph{Proc. {ACM} Program. Lang.}}
  \bibinfo{volume}{6}, \bibinfo{number}{{ICFP}} (\bibinfo{year}{2022}),
  \bibinfo{pages}{540--569}.
\newblock
\urldef\tempurl%
\url{https://doi.org/10.1145/3547641}
\showDOI{\tempurl}


\bibitem[Martin{-}L{\"{o}}f(1984)]%
        {DBLP:books/daglib/0000395}
\bibfield{author}{\bibinfo{person}{Per Martin{-}L{\"{o}}f}.}
  \bibinfo{year}{1984}\natexlab{}.
\newblock \bibinfo{booktitle}{\emph{Intuitionistic type theory}}.
  \bibinfo{series}{Studies in proof theory}, Vol.~\bibinfo{volume}{1}.
\newblock \bibinfo{publisher}{Bibliopolis}.
\newblock
\showISBNx{978-88-7088-228-5}


\bibitem[Mitchell and Moggi(1991)]%
        {DBLP:journals/apal/MitchellM91}
\bibfield{author}{\bibinfo{person}{John~C. Mitchell} {and}
  \bibinfo{person}{Eugenio Moggi}.} \bibinfo{year}{1991}\natexlab{}.
\newblock \showarticletitle{Kripke-Style Models for Typed lambda Calculus}.
\newblock \bibinfo{journal}{\emph{Ann. Pure Appl. Log.}} \bibinfo{volume}{51},
  \bibinfo{number}{1-2} (\bibinfo{year}{1991}), \bibinfo{pages}{99--124}.
\newblock
\urldef\tempurl%
\url{https://doi.org/10.1016/0168-0072(91)90067-V}
\showDOI{\tempurl}


\bibitem[Norell(2008)]%
        {DBLP:conf/afp/Norell08}
\bibfield{author}{\bibinfo{person}{Ulf Norell}.}
  \bibinfo{year}{2008}\natexlab{}.
\newblock \showarticletitle{Dependently Typed Programming in Agda}. In
  \bibinfo{booktitle}{\emph{Advanced Functional Programming, 6th International
  School, {AFP} 2008, Heijen, The Netherlands, May 2008, Revised Lectures}}
  \emph{(\bibinfo{series}{Lecture Notes in Computer Science},
  Vol.~\bibinfo{volume}{5832})}, \bibfield{editor}{\bibinfo{person}{Pieter
  W.~M. Koopman}, \bibinfo{person}{Rinus Plasmeijer}, {and}
  \bibinfo{person}{S.~Doaitse Swierstra}} (Eds.).
  \bibinfo{publisher}{Springer}, \bibinfo{pages}{230--266}.
\newblock
\urldef\tempurl%
\url{https://doi.org/10.1007/978-3-642-04652-0_5}
\showDOI{\tempurl}


\bibitem[Paykin et~al\mbox{.}(2017)]%
        {DBLP:conf/popl/Paykin0Z17}
\bibfield{author}{\bibinfo{person}{Jennifer Paykin}, \bibinfo{person}{Robert
  Rand}, {and} \bibinfo{person}{Steve Zdancewic}.}
  \bibinfo{year}{2017}\natexlab{}.
\newblock \showarticletitle{{QWIRE:} a core language for quantum circuits}. In
  \bibinfo{booktitle}{\emph{Proceedings of the 44th {ACM} {SIGPLAN} Symposium
  on Principles of Programming Languages, {POPL} 2017, Paris, France, January
  18-20, 2017}}, \bibfield{editor}{\bibinfo{person}{Giuseppe Castagna} {and}
  \bibinfo{person}{Andrew~D. Gordon}} (Eds.). \bibinfo{publisher}{{ACM}},
  \bibinfo{pages}{846--858}.
\newblock
\urldef\tempurl%
\url{https://doi.org/10.1145/3009837.3009894}
\showDOI{\tempurl}


\bibitem[Rennela and Staton(2020)]%
        {DBLP:journals/lmcs/RennelaS19}
\bibfield{author}{\bibinfo{person}{Mathys Rennela} {and} \bibinfo{person}{Sam
  Staton}.} \bibinfo{year}{2020}\natexlab{}.
\newblock \showarticletitle{Classical Control, Quantum Circuits and Linear
  Logic in Enriched Category Theory}.
\newblock \bibinfo{journal}{\emph{Log. Methods Comput. Sci.}}
  \bibinfo{volume}{16}, \bibinfo{number}{1} (\bibinfo{year}{2020}).
\newblock
\urldef\tempurl%
\url{https://doi.org/10.23638/LMCS-16(1:30)2020}
\showDOI{\tempurl}


\bibitem[Scholz(2014)]%
        {DBLP:conf/pepm/Scholz14}
\bibfield{author}{\bibinfo{person}{Sven{-}Bodo Scholz}.}
  \bibinfo{year}{2014}\natexlab{}.
\newblock \showarticletitle{Partial evaluation as universal compiler tool:
  experiences from the {SAC} Eco system}. In
  \bibinfo{booktitle}{\emph{Proceedings of the {ACM} {SIGPLAN} 2014 workshop on
  Partial evaluation and program manipulation, {PEPM} 2014, January 20-21,
  2014, San Diego, California, {USA}}},
  \bibfield{editor}{\bibinfo{person}{Wei{-}Ngan Chin} {and}
  \bibinfo{person}{Jurriaan Hage}} (Eds.). \bibinfo{publisher}{{ACM}},
  \bibinfo{pages}{95--96}.
\newblock
\urldef\tempurl%
\url{https://doi.org/10.1145/2543728.2543747}
\showDOI{\tempurl}


\bibitem[Wood and Atkey(2022)]%
        {DBLP:conf/esop/WoodA22}
\bibfield{author}{\bibinfo{person}{James Wood} {and} \bibinfo{person}{Robert
  Atkey}.} \bibinfo{year}{2022}\natexlab{}.
\newblock \showarticletitle{A Framework for Substructural Type Systems}. In
  \bibinfo{booktitle}{\emph{Programming Languages and Systems - 31st European
  Symposium on Programming, {ESOP} 2022, Held as Part of the European Joint
  Conferences on Theory and Practice of Software, {ETAPS} 2022, Munich,
  Germany, April 2-7, 2022, Proceedings}} \emph{(\bibinfo{series}{Lecture Notes
  in Computer Science}, Vol.~\bibinfo{volume}{13240})},
  \bibfield{editor}{\bibinfo{person}{Ilya Sergey}} (Ed.).
  \bibinfo{publisher}{Springer}, \bibinfo{pages}{376--402}.
\newblock
\urldef\tempurl%
\url{https://doi.org/10.1007/978-3-030-99336-8\_14}
\showDOI{\tempurl}


\end{thebibliography}

\end{document}